\documentclass[%
 reprint,
 amsmath,amssymb,
 aps,
 superscriptaddress
]{revtex4-2}
\usepackage{graphicx}
\usepackage{dcolumn}
\usepackage{bm}
\usepackage{amssymb}
\begin{document}
\preprint{APS/123-QED}

\title{Line defects in nematic liquid crystals as charged superelastic rods with negative twist--stretch coupling}
\author{Shengzhu Yi}
\affiliation{Department of Physics, The Hong Kong University of Science and Technology, Clear Water Bay, Kowloon, Hong Kong SAR}
\affiliation{Department of Mechanical and Energy Engineering, Southern University of Science and Technology, Shenzhen, Guangdong, China}
\author{Hao Chen}
\affiliation{Department of Mechanical and Energy Engineering, Southern University of Science and Technology, Shenzhen, Guangdong, China}
\author{Xinyu Wang}
\affiliation{Department of Physics, The Hong Kong University of Science and Technology, Clear Water Bay, Kowloon, Hong Kong SAR}
\author{Miao Jiang}
\affiliation{Department of Mechanical and Energy Engineering, Southern University of Science and Technology, Shenzhen, Guangdong, China}
\author{Bo Li}
\affiliation{Institute of Biomechanics and Medical Engineering, Applied Mechanics Laboratory, Department of Engineering Mechanics, Tsinghua University, Beijing 100084, China}
\author{Qi-huo Wei}
\email{weiqh@sustech.edu.cn}
\affiliation{Department of Mechanical and Energy Engineering, Southern University of Science and Technology, Shenzhen, Guangdong, China}
\affiliation{Center for Complex Flows and Soft Matter Research, Southern University of Science and Technology, Shenzhen 518055, China}
\author{Rui Zhang}
\email{ruizhang@ust.hk}
\affiliation{Department of Physics, The Hong Kong University of Science and Technology, Clear Water Bay, Kowloon, Hong Kong SAR}

\date{\today}
\setlength{\parskip}{0em}
\begin{abstract}
Topological defects are a ubiquitous phenomenon in diverse physical systems. In nematic liquid crystals (LCs), they are dynamic, physicochemically distinct, sensitive to stimuli, and are thereby promising for a range of applications. However, our current understanding of the mechanics and dynamics of defects in nematic LCs remain limited and are often overwhelmed by the intricate details of the specific systems.
Here, we unify singular and nonsingular line defects as superelastic rods and combine theory, simulation, and experiment to quantitatively measure their effective elastic moduli, including line tension, torsional rigidity, and twist--stretch coefficient. Interestingly, we found that line defects exhibit a negative twist--stretch coupling, meaning that twisted line defects tend to unwind under stretching, which is reminiscent of DNA molecules. A patterned nematic cell experiment further confirmed the above findings. Taken together, we have established an effective elasticity theory for nematic defects, paving the way towards understanding and engineering their deformation and transformation in driven and active nematic materials.
\end{abstract}
\maketitle

\textit{Introduction-} Topological defects are ubiquitous in nature, such as various condensed matter (e.g., crystals, liquid crystals, superconductors, and superfluids), optics, early universe, and particle physics \cite{wittmannParticleresolvedTopologicalDefects2021,senyukTopologicalColloids2013,poyInteractionCoassemblyOptical2022}. 
In materials science, topological defects are pivotal for dictating the properties and functionalities of a material \cite{baggioliTopologicalDefectsReveal2023}.
Liquid crystals (LCs) consisting of anisotropic molecules exhibit phases intermediate between fluids and solids \cite{gennesPhysicsLiquidCrystals1993}. Recent advances in experimental control of molecule orientations such as photo alignment technique have enabled creation and manipulation of defect structures \cite{guoPhotopatternedDesignerDisclination2021,guoHighResolutionHighThroughputPlasmonic2016,nikkhouLightcontrolledTopologicalCharge2015}. Therefore, a unique combination of controllability and fluidity of LCs makes it an ideal model system onto which to study the universal physics of defects.

In nematic LCs, topological defects can be either singular or non-singular \cite{davidsonChiralStructuresDefects2015,polakOpticalDeterminationSaddlesplay1994}, and are also named ``disclinations''. They are ultra-sensitive to the environment, and are thereby promising for applications in biosensors \cite{sadatiLiquidCrystalEnabled2015,linEndotoxinInducedStructuralTransformations2011}, photonics \cite{martinez-gonzalezDirectedSelfassemblyLiquid2017}, and self-assembly \cite{yoshidaThreedimensionalPositioningControl2015,jiangCollectiveTransportReconfigurable2023,wangExperimentalInsightsNanostructure2016,rahimiSegregationLiquidCrystal2017,wangTopologicalDefectsLiquid2016}. In active nematic LCs, the autonomous motion of defects offers new opportunities in microfluidics for tailoring flow patterns, manipulating material transport, and executing logic operations  \cite{wangInterplayActiveStress2021,zhangDynamicStructureActive2016,zhangAutonomousMaterialsSystems2021}. Moreover, defects can also be programmed into LC elastomers, yielding well-defined, nontrivial deformation modes when exposed to certain stimulus \cite{wareVoxelatedLiquidCrystal2015,guinLayeredLiquidCrystal2018}. Therefore, topological defects confer novel features to LC systems, and are useful for a range of emerging applications. This creates a strong need for understanding their viscoelastic properties and engineering their motion and deformation.

Disclinations carry a line tension, which is firstly introduced by P. G. de Gennes \cite{gennesPhysicsLiquidCrystals1993} and further theoretically considered by M. Kleman \cite{cladisNonsingularDisclinationsStrength1972} and S. Zumer \cite{polakOpticalDeterminationSaddlesplay1994}. Laser tweezer technique applied by Smalyukh and coauthors has provided an intuitive and feasible method to change the local director field and to stretch disclinations, enabling the estimate of their line tension \cite{smalyukhOpticalTrappingManipulation2006,petit-garridoHealingDefectsInterface2011,smalyukhOpticalTrappingColloidal2005}.
Other works demonstrated that line defects can be deformed by applying hydrodynamic flows \cite{gettelfingerFlowInducedDeformation2010, coparMicrofluidicControlTopological2020} or external fields \cite{fukudaFieldinducedDynamicsStructures2013,harkaiElectricFieldDriven2020}. In addition, recent efforts in controlling disclinations have been focused on using light or surfactant molecules to alter surface anchoring conditions \cite{jiangActiveTransformationsTopological2022,senyukTransformationElasticDipoles2021}. Despite the above-mentioned rich parameters to manipulate disclinations, our current understanding of their mechanical response to stimuli remains elusive. Existing models often rely on time-consuming calculations or are developed for specific defect types or structures \cite{schimmingKinematicsDynamicsDisclination2023,ostermanRelaxationKineticsStretched2010,longFrankReadMechanismNematic2022,modinTunableThreedimensionalArchitecture2023,longGeometryMechanicsDisclination2021}.
This limits our capacity to design topological defects with desirable morphogenesis and dynamics for their potential applications.

Recent advances in photo-patterning technique have enabled precise design of arbitrary shaped disclinations in nematic cells by imprinting a specific anchoring pattern on the substrates \cite{mengTopologicalSteeringLight2023}.
There is a recent interest in combining this technique with straight-forward mechanical methods, e.g., twisting, to generate versatile defect structures \cite{MoireEffectEnablesVersatile}.
In this work, we extend this technique to create line defects, and conduct mechanical tests on them to quantitatively measure their elastic moduli, including line tension, torsional rigidity, bending modulus, and twist--stretch coupling coefficient. Our work shows that line defects, either singular or non-singular, can be mapped to charged superelastic rods. Interestingly, our theory, simulation, and experiment have found that line defects exhibit a negative twist--stretch coupling, implying that they tend to bend when subjected to stretch and twist simultaneously. We further employ our effective elasticity theory of disclinations to design nontrivial defect structures.
As such, we have developed a comprehensive elasticity theory for line defects in nematic LCs, providing a convenient framework to understand and engineer their intricate mechanical responses to stimuli.

\begin{figure}[t]
\includegraphics{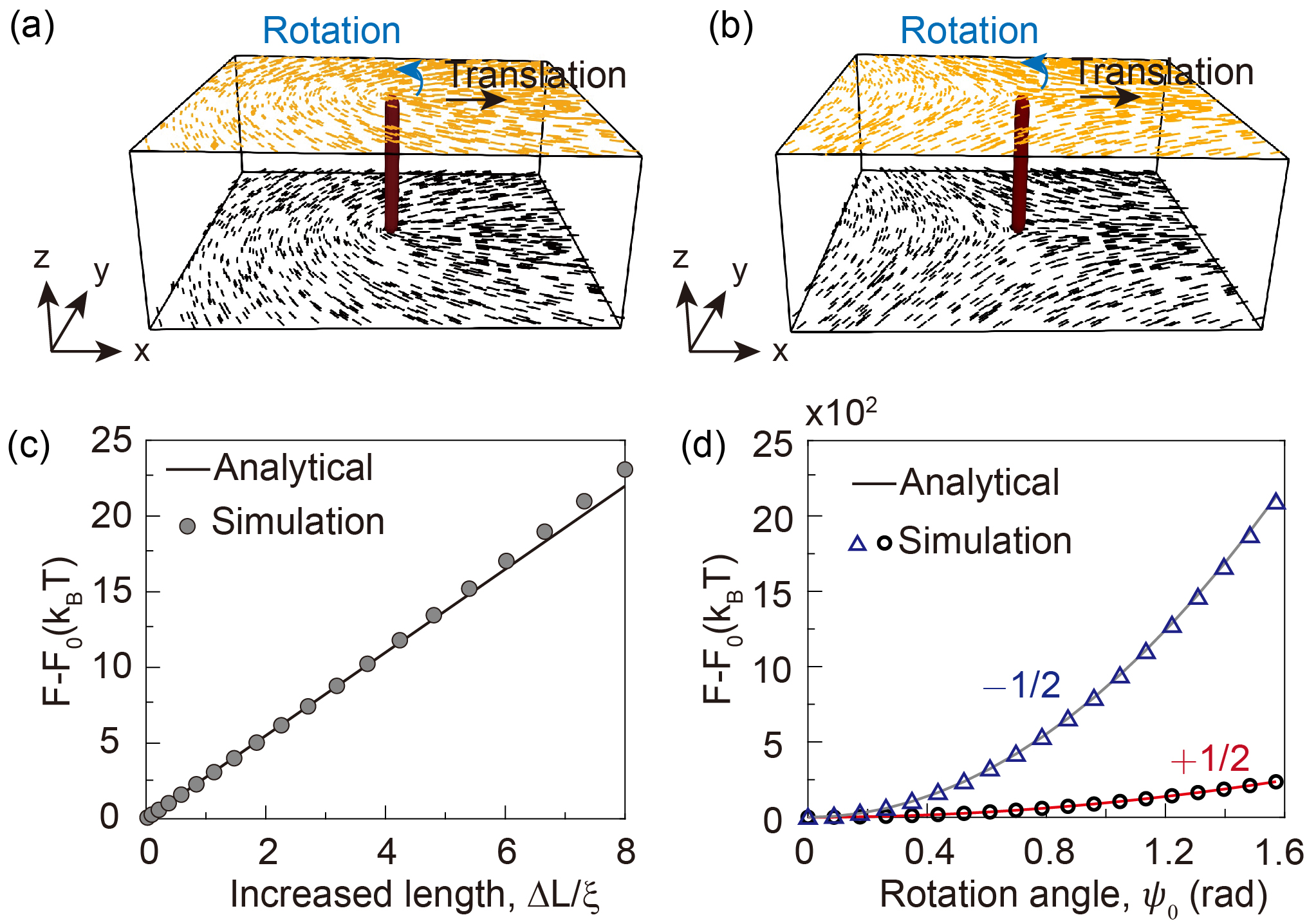}
\caption{\label{fig:one} Measurements of Line tension and torsional rigidity of a disclination. Simulation boxes of a $+1/2$ (a) and a $-1/2$ (b) disclination.  As indicated by the arrows, the top surface was translated or rotated to deform the disclination. The increased free energy versus the change of the defect length $\Delta L$ (c) and the rotation angle $\psi_0$ (d). Markers and solid lines represent simulated and theoretical results, respectively.}
\end{figure}

\textit{Results-} Our analysis begins with an examination of the line tension of singular defects. A disclination with topological charge \textit{m} can be generated in a nematic cell by patterning a point defect of strength \textit{m} on both top and bottom substrates [Fig.~\ref{fig:one}(a),~\ref{fig:one}(b)]. The charge \textit{m} corresponds to the rotation of the director when one circumnavigates the defect core \cite{shamsTheoryModelingNematic2012}. Initially, the two surface patterns are well aligned such that the line defect is normal to the substrates and its winding is invariant along the line. Stretching of the defect can be realized by laterally translating the top surface by a distance while fixing the bottom surface (Fig. S1 and Movie 1~\cite{Supple}). Under the one-elastic-constant approximation, the elastic free energy penalty due to orientational distortions can be described by \cite{sidkySilicoMeasurementElastic2018}: $\textit{F}=\frac{1}{2}\int_V\textit{K}(\triangledown|\textbf{n}|)^2\mathrm{d}V$, where \textit{K} is the Frank elastic modulus. By rewriting the increase of the elastic energy as a function of the increased length of the defect upon stretching, one can obtain the line tension $\tau_{s}$ of the singular disclination (Supplementary Sec. \uppercase\expandafter{\romannumeral1}~\cite{Supple}):
\begin{equation}\label{eq:onee}
\tau_{s}=\pi Km^2\ln\left(\frac{\textit{R}}{\textit{r}}\right)+\textit{f}_c,
\end{equation}
where \textit{r} and $f_c$ represent the radius and free energy cost per length of the defect core, respectively, and $R$ denotes the radius of the lateral integration volume representing the size of the system. Eq.~(\ref{eq:onee}) is further confirmed in our continuum simulations based on Landau--de Gennes free energy functional (Fig.~\ref{fig:one}(c)). 
Note that the line tension of half-integer disclinations is typically several tens of pN~\cite{smalyukhOpticalTrappingColloidal2005}, 
which is one order of magnitude smaller than the stretch modulus $S=1100\pm200$ pN of DNA molecules \cite{goreDNAOverwindsWhen2006}.

To further the analogy between line defects and superelastic rods, we next consider the torsional rigidity of disclinations. We apply a twist to a disclination by rotating the top surface with respect to the defect core by an angle $\psi_0$ (Movie 2~\cite{Supple}). The increased free energy associated with this distortion is $\Delta F=\frac{1}{2}KA(1-m)^2\psi_0^2/\textit{L}_z$, where $A=\pi \textit{R}^2$ is the lateral surface area of the integration volume. There is no energy penalty for a disclination of strength $m=+1$, as it has rotational symmetry. By comparing the above equation to the elastic energy penalty of an elastic rod under a twist deformation \cite{bergouDiscreteElasticRods2008} $F_t=\frac{1}{2}C_t\psi_0^2/L$, we find the torsional rigidity $C_t$ of a disclination to be (Supplementary Sec. \uppercase\expandafter{\romannumeral2}~\cite{Supple}): $C_t=KA(1-m)^2$. Compared to the line tension, which depends on the system size logarithmically, $C_t$ is proportional to $R^2$, and is therefore more sensitive to the system size. The above equation also implies that negative-charge disclinations are harder to twist than positive-charge disclinations with the same magnitude of $|m|$ (Fig.~\ref{fig:one}(d)). For example, despite that $\pm1/2$ disclinations share the same line tension, the torsional rigidity of $-1/2$ disclinations is 9 times that of $+1/2$ disclinations (Fig.~S2).
Note that even a uniform nematic exhibits a torsional rigidity 
$C_t^\text{uni}=KA$ as the twisting operation can cause elastic energy penalty, and interestingly, $+1/2$ and $+3/2$ disclinations carry a smaller $C_t$.

Next, we investigate the behavior of a disclination under a combination of twist and stretch. Specifically, we apply a displacement $(\Delta x,\Delta y)=(\lambda_1^0,\lambda_2^0 )$ to the top surface, followed by a rotation of angle $\psi_0$ with respect to the defect center. Using Frank elasticity theory, we find that the increased free energy takes the following form (Supplementary Sec. \uppercase\expandafter{\romannumeral3}~\cite{Supple}):
\begin{figure}[t]
\includegraphics{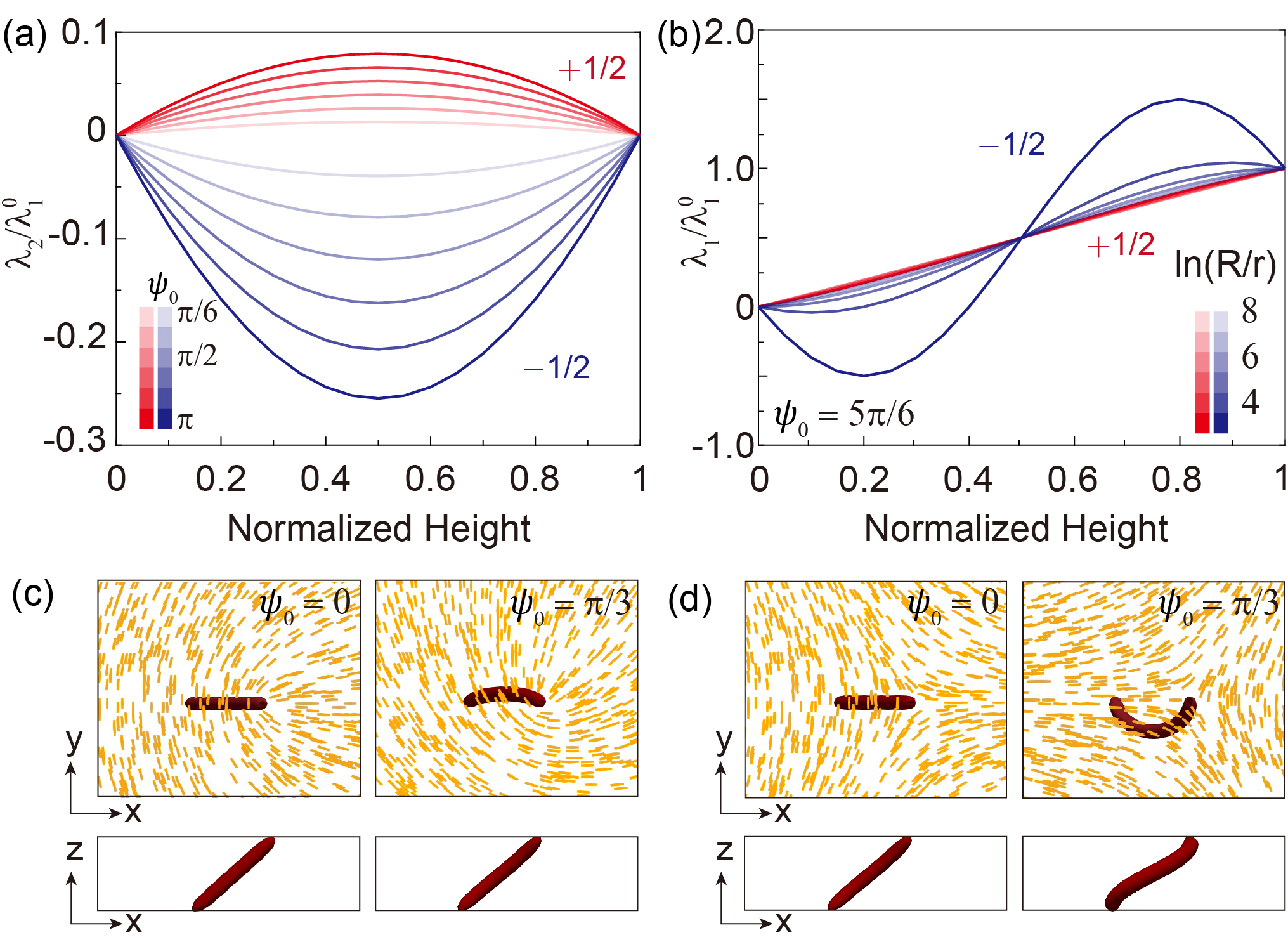}
\caption{\label{fig:two}Configurations of disclinations under stretch and twist. The profiles of $+1/2$ (red lines) and $-1/2$ (blue lines) disclinations depend on the twist angle $\psi_0$ (a) and the logarithm term $\ln\left(\frac{R}{r}\right)$ (b). $\lambda_1$ and $\lambda_2$ represent the displacement of defects in the $x$- and $y$-direction, respectively. The above curves are calculated using boundary conditions $\lambda_1(L_z)=\lambda_1^0, \lambda_2(L_z)=0$. Simulated results of stretched and twisted disclinations with winding number of $+1/2$ (c) and $-1/2$ (d). Directors on the top surface are indicated by the yellow lines.}
\end{figure}
\begin{eqnarray}
\Delta F= &&\frac{1}{2}\pi Km^2\ln\left(\frac{R}{r}\right)\frac{(\lambda^0)^2}{L_z}+\frac{1}{2}\pi KR^2(1-m)^2\frac{\psi_0^2}{L_z}\nonumber\\
& &
-\frac{1}{3}\pi K(1-m)^2\frac{(\lambda^0)^2\psi_0^2}{L_z\ln\left(\frac{R}{r}\right)},%
\label{eq:three}
\end{eqnarray}
where $\lambda^0=\sqrt{(\lambda_1^0)^2+(\lambda_2^0)^2}$ is the translation distance of the top surface. The first and second term in the above refer to the elastic energy caused by stretch and twist, respectively, which have been discussed in the above. The third term corresponds to a decrease in the free energy via a non-positive twist--stretch coupling. This new term emerges only when stretching $(\lambda^0\neq0)$ and twisting $(\psi_0\neq0)$ deformation are both present, and is akin to the twist--stretch coupling term for DNA molecules~\cite{goreDNAOverwindsWhen2006}. Therefore, we expect that twisted disclinations should behave like DNA molecules, which can overwind under stretching. Note for DNA molecules, the twist--stretch coupling term depends linearly on $\lambda^0$ and $\psi_0$. Whereas for nematic disclinations, this term depends quadratically on $\lambda^0$ and $\psi_0$, and is thereby weaker than that of DNA molecules.

To solve the defect shape without loss of generality, we consider a slight stretch of the twisted disclination along the $x$ direction, i.e., $\lambda_1^0\neq0$, $\lambda_2^0=0$, and $\lambda_1^0\ll R$. Our theory shows that the disclination profile can be approximated by a parabolic shape within a plane perpendicular to the shear ($xz$) plane [Fig.~\ref{fig:two}(a)], and the bend direction depends on the sign of $m$ (Supplementary Sec. \uppercase\expandafter{\romannumeral3}~\cite{Supple}). The degree of bending, which can be characterized by $d$, the maximum displacement projected on the $xy$ plane, is found linearly dependent on the twist angle and the stretch length. Our simulations confirm that $\pm1/2$ disclinations indeed bend in opposite directions with a magnitude dependent on the winding number, the translation distance, and the twist angle (Movie 3~\cite{Supple}). At a large twist angle and small $\ln\left(\frac{R}{r}\right)$, the disclination takes a sinusoidal shape (Fig.~\ref{fig:two}(b)). Note that the twist--stretch coupling term is approximately two to three orders of magnitude lower than the energy penalty caused by translation or twist, its influence on the disclination shape is significant.
\begin{figure}[b]
\includegraphics{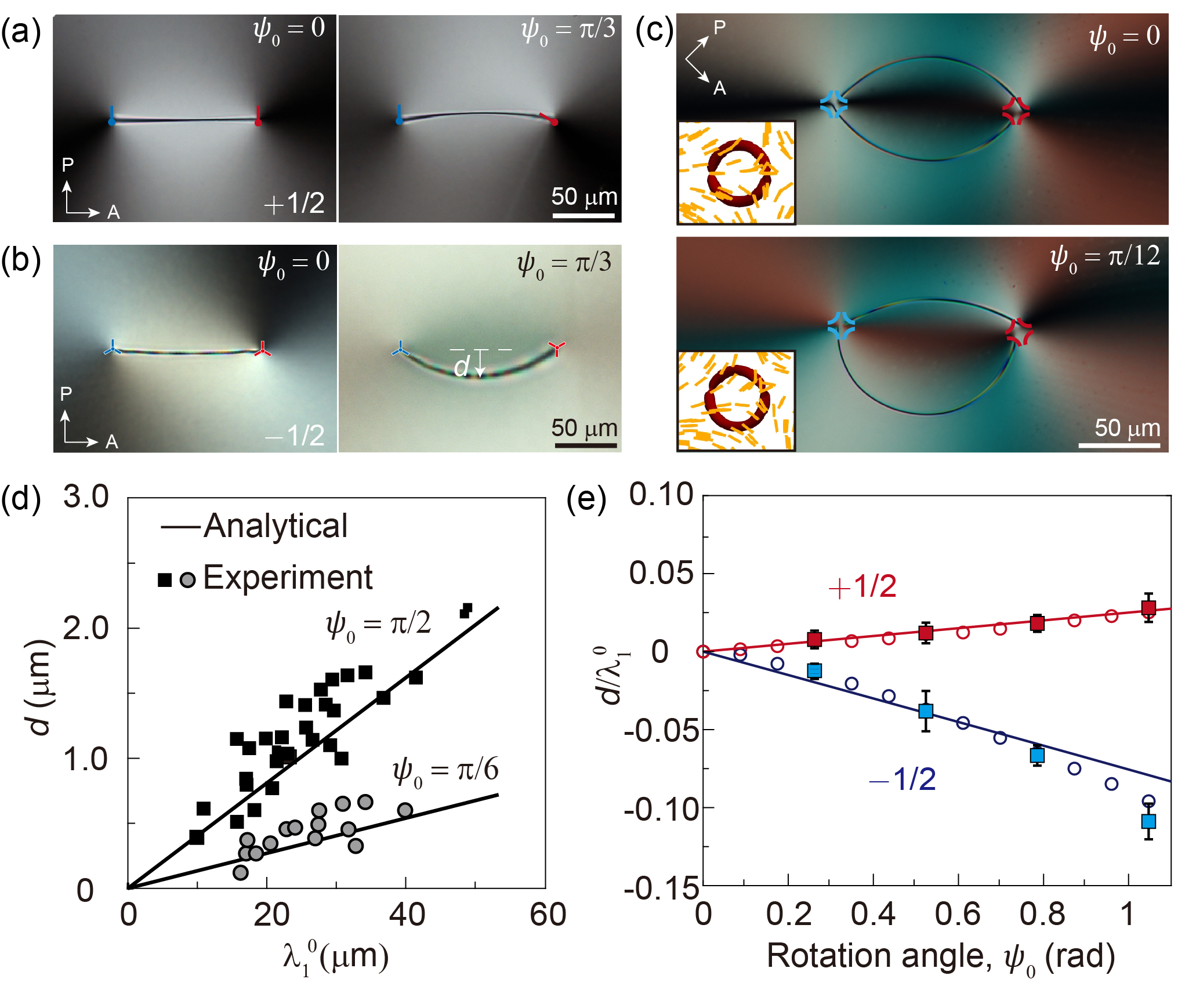}
\caption{\label{fig:three} Experimental results of $+1/2$ (a) and $-1/2$ (b) disclinations subjected to stretch and twist. (c) Simulated (insets) and experimental shapes of stretched defects $(m=-1)$ under various twist angles. (d) Maximum displacement of $+1/2$ disclinations versus the translation distance $\lambda_1^0$. (e) The deflection of $\pm1/2$ disclinations depends on the twist angle $\psi_0$. Error bars indicate standard deviation of the measured data. Squares, circles, and lines represent experimental, simulated, and theoretical results, respectively.}
\end{figure}

To test our finding of negative twist--stretch coupling, we performed experiments based on photo-patterning method. The cell was constructed by two patterns with identical defects but different orientations and a horizontal displacement $\lambda^0$. As expected, the resultant disclination exhibits a bent shape when viewed from top, and the bent direction is opposite between positive and negative defects, confirming our predictions [Fig.~\ref{fig:three}(a),~\ref{fig:three}(b)  and Fig. S3~\cite{Supple}]. We further measure the degree of bending, $d/\lambda_1^0$, as a function of the twist angle $\psi_0$, and find quantitative agreement between the experiment and the predictions [Fig.~\ref{fig:three}(e)]. In addition, we demonstrate that highly nontrivial topological structures can be achieved by our effective elasticity theory [Fig.~\ref{fig:three}(c)]. A disclination of strength $-1$ will split into two disclinations of strength $-1/2$ in the bulk. Before twist, simulations and experiments show that the two line defects form a symmetric spindle-like shape due to a competition between their repulsive forces and line tensions. When a twist is applied, the twist--stretch coupling breaks the symmetry and enables the formation of different bent shapes of disclinations depending on the twist angle (Movie 4~\cite{Supple}). 

\begin{figure}[t]
\includegraphics{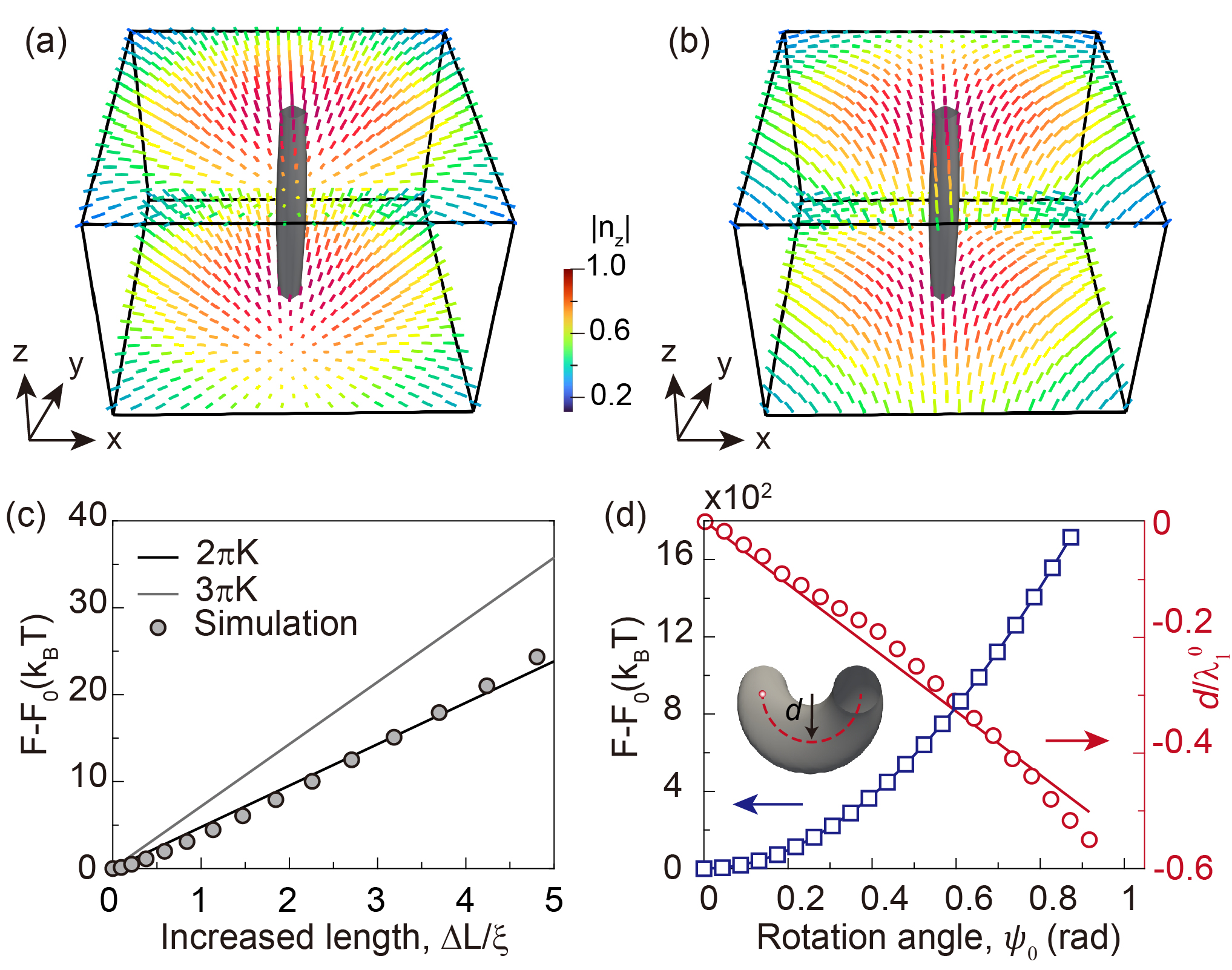}
\caption{\label{fig:four}Elastic properties of non-singular unit-strength defects. Simulation boxes of a  $+1$ (a) and a $-1$ (b) nonsingular disclination. (c) Free energy depending on the increased length $\Delta L$. (d) Twist penalties and maximum displacement depending on the twist angle $\psi_0$.The non-singular defects are visualized by an isosurface $|n_z|=0.995$. The insert shows the bent  $-1$ non-singular disclination caused by twist--stretch coupling.}
\end{figure}

We further extend our analysis to non-singular defects. For this type of defects, the directors escape into the third dimension and their cores remain nematic. Here we focus on non-singular defects of strength of $m=\pm 1$ [Fig.~\ref{fig:four}(a),~\ref{fig:four}(b)].  An isosurface of $|n_z|=0.995$ is plotted to visualize their `cores'. They become elongated under stretch and the increased free energy has a linear dependence on their length. Assuming $K=K_{11}=K_{22}=K_{33}$, the line tension of a non-singular defects of strength $m$ becomes (Supplementary Sec. \uppercase\expandafter{\romannumeral4}~\cite{Supple}): 
\begin{equation}
\tau_{n}=(1+m+m^2)\pi K -m\pi K_{24}.%
\label{eq:four}
\end{equation}
In the one-constant approximation $(K_{24}=K)$, the line tension of disclinaitons with $|\textit{m}|=1$ becomes $\tau=2\pi K$, as confirmed in our simulation [Fig.~\ref{fig:four}(c)].
This explains why non-singular $+1$ defects prevail in materials with high values of $K_{24}$, such as nematic lyotropic chromonic liquid crystals \cite{crawfordSurfaceElasticMolecularanchoring1992}.

The torsional rigidity of non-singular defects can be approximately calculated as (Supplementary Sec. \uppercase\expandafter{\romannumeral5}~\cite{Supple}) $C_t\approx KA(4\ln\left(2\right)-2)(1-m)^2$, slightly smaller than that of singular defects. This is consistent with the fact that twist penalties can be released when the planar directors escape to the third dimension. Again, non-singular defects with strength of $+1$ have vanishing twist modulus. Non-singular defects also exhibits twist--stretch coupling properties [Fig.~\ref{fig:four}(d) and Movie 5~\cite{Supple}] and their deformation is similar to singular disclinations. There is a quantitative difference in the degree of deformation. The bending magnitude $d/\lambda_1^0\psi_0\sim-0.58$ for a non-singular $-1$ defect is considerably larger than that of a singular $-1$ disclination $\sim-0.12$. Nevertheless, our elasticity theory can successfully unify the two types of line defects.

Line defects can also be bent via defect--defect interactions. When two line defects of opposite charge are close to each other, they bend toward each other. This phenomenon can be explained through the interplay between their elastic attractions and the bending energy, similar to the mechanical behavior observed in two charged elastic beams with opposite charge signs (Fig. S4~\cite{Supple}). The bending modulus $C_b$ of the ``elastic beam'' is a direct consequence of its line tension, one can find their dependence to be: $C_b=\tau_s L_z^2/12$ (Supplementary Sec. \uppercase\expandafter{\romannumeral5} ~\cite{Supple}). This elastic modulus is also size dependant. Also note that the Poisson's ratio of disclinations is effectively 0, as their core radii are unchanged during a longitudinal stretch (Fig. S5~\cite{Supple}). Since the chemical components of defects are identical to their surroundings, the elongation of disclinations can be achieved by converting nearby molecules in the nematic phase into the isotropic core, without changing their lateral size.

\textit{Discussion-} Coarse-grained models of topological defects can be a powerful tool for understanding and controlling their dynamics and morphogensis. In two-dimensional active nematics, for example, Coulomb gas model has been successful in elucidating the dynamics of self-propelled point defects \cite{shankarHydrodynamicsActiveDefects2019}. In this work, we extend the concept of line tension of nematic defects to examining their other mechanical constants, including torsional rigidity, bending modulus, and twist--stretch coupling coefficient. Combining photo-patterning technique with the straightforward mechanical method, we are able to quantitatively measure these constants in both simulations and experiments, which give rise to quantitatively consistent results, demonstrating the validity of the superelastic-rod framework.
However, the effective elastic moduli, including line tension and torsional rigidity, are system size dependent, except that the line tension of non-singular defects depends on $K_{24}$. This dependence may lead to novel ways of measuring $K_{24}$. 

Remarkably, our theory and simulation predict a non-positive twist--stretch coupling for disclinations. This coupling is responsible for the bent shape of the disclination under a combination of stretch and twist. 
Note that the bending of disclinations can also be explained by an effective Peach--Koehler force \cite{longFrankReadMechanismNematic2022} or a Lorentz-like force \cite{modinTunableThreedimensionalArchitecture2023}. Here, our twist--stretch coupling theory can precisely capture the bent shape of the disclination and is independent of any specifics of a system. There is also an excellent agreement between analytical theory, continuum simulation, and experiment in terms of the deformed structures of the disclination. Taken together, our work establishes a striking analog between disclinaitons and charged superelastic rods, allowing an intuitive and quantitative understanding of the mechanics of disclinations, and facilitating their emerging applications such as design of autonomous materials, dynamic self-assembly, and material transport. 
\setlength{\parskip}{1em}

\textit{Acknowledgement-} Q. W. acknowledges the financial support by Shenzhen Science and Technology Innovation Committee GJHZ20200731095212036 and National Natural Science Foundation of China No.12174177. R. Z. acknowledges the financial support from Hong Kong Research Grants Council through No. 16300221 and No. 26302320. R. Z. thanks Robin Selinger and Jonathan Selinger for the helpful discussions.
Shengzhu Yi and Hao Chen contributed equally to this work.

\bibliography{SuperelasticRods}
\end{document}